\documentstyle[12pt]{article}
\begin{document}

\topmargin 0pt

\oddsidemargin -3.5mm

\headheight 0pt

\topskip 0mm
\addtolength{\baselineskip}{0.20\baselineskip}
\begin{flushright}
SNUTP-98-117 \\
SOGANG-HEP 250/98 \\
hep-th/9811033
\end{flushright}
\vspace{0.3cm}
\begin{center}
    {\large \bf  Symmetry Algebras in
      Chern-Simons Theories with Boundary: Canonical Approach }  
\end{center}
\vspace{0.3cm}
\begin{center}
 Mu-In Park\footnote{Electronic address: mipark@physics.sogang.ac.kr }
\footnote{ On leave from {\it Center for
  Theoretical Physics, Massachusetts Institute of Technology,
  Cambridge, Massachusetts 02139 U.S.A.}}\\
{Center for Theoretical Physics, 
Seoul National University, \\
Seoul, 151-742, Korea} \\
{\it and} \\
{Basic Science Research Institute, Sogang University,\\
C. P. O. Box 1142, Seoul 100-611, Korea}\\
\end{center}
\vspace{0.3cm}
\begin{center}
    {\bf ABSTRACT}
\end{center}
I consider the classical Kac-Moody algebra and Virasoro algebra in
Chern-Simons theory with boundary within the Dirac's canonical method
and Noether procedure. It is shown that the usual (bulk) Gauss law
constraint becomes a second-class constraint because of the boundary
effect. From this fact, the Dirac bracket can be constructed explicitly without
introducing additional gauge conditions and the classical Kac-Moody
and Virasoro algebras are obtained within the usual
Dirac method. The equivalence to the symplectic reduction method is
presented and the connection to the Ba\~nados's work is clarified.
It is also considered the generalization to the Yang-Mills-Chern-Simons
theory where the diffeomorphism symmetry is broken by the
(three-dimensional) Yang-Mills 
term. In this case, the same Kac-Moody algebras are obtained
although the two theories are sharply different in the canonical
structures. The both models realize the holography
principle explicitly and the pure CS theory reveals the correspondence
of the {\it Chern-Simons theory with boundary/conformal field theory},
which is more fundamental 
and generalizes the conjectured {\it anti-de Sitter/conformal field theory }
correspondence.

\vspace{0.3cm}
\begin{flushleft}

November 1998 \\
\end{flushleft}

\newpage

\begin{center}
{\bf I. Introduction}
\end{center}

Recently, there has been vast interest on the role of the space boundary in
diverse areas of physics [1 - 3]. Though the complete understanding of
the boundary physics has not been attained yet, this boundary theory
opened new rich areas both in physics and 
mathematics. One of the interesting areas is what comes from the existence of
the central terms in the Kac-Moody algebra and Virasoro algebra even
at ``classical'' level. These unusual classical algebras were found
first in the asymptotic isometry group $SO(2,2)$ of the
three-dimensional anti-de Sitter
space $(AdS_{2+1})$ more than 10 years ago [4]. It is only in recent times that
this algebra was applied to a practical physical problem of the
statistical entropy calculation for $BTZ$ black hole [5] by
Strominger, which might provide important clues for understanding
the mystery of black holes [6].

On the other hand, recently there was also an interesting report on the
similar ``classical'' central terms in the Kac-Moody and Virasoro algebras
in the Chern-Simons (CS) theory [7] with boundary [1,8] using the
Regge-Teitelboim's canonical method [4,9] by Ba\~nados [10]. 
However, in this work he considered several hypothetical procedures
which make it difficult to understand the work by the usual and
familiar field theory methods. Motivated by this problem, the
well-known symplectic reduction method [11] was considered more recently
[12] and the Ba\~nados's Kac-Moody and Virasoro algebras were {\it
  rigorously} derived with the 
help of the Noether procedure for constructing the conserved charges [13].
Then, following the equivalence of the CS theory
to the (2+1)-dimensional gravity theory with a
cosmological constant [14, 15], it was straightforward to apply the
Ba\~nados's algebras to the BTZ black hole (negative cosmological constant)
entropy [16] and Kerr-de Sitter space (positive cosmological constant)
entropy [17] $a'~ la$ Strominger.\footnote{ Alternative approaches have
  been considered by 
  Maldacena-Strominger and Ba\~nados-Brotz-Ortiz for the de-Sitter
  space [18].} A merit of this approach was that the Virasoro algebra
can be found for any radii, although the details of the 
central charge depend on the boundary diffeomorphism
({\it Diff}). However, even in this symplectic reduction method, where
only the boundary degrees of freedom are treated by imposing the
(Gauss law) constraint, the origin
of the center was not understood at the more fundamental level as the
result of interaction of bulk and boundary degrees of freedom.
Moreover, the connection to the Ba\~nados's work [9] was not
clear either.

In this paper, I clarify these issues within the usual Dirac
method [19]. In this method, several remarkable implications of the
classical center are manifest. In Sec. II, it is shown that 
the usual (bulk) Gauss law constraint of the (pure) CS theory becomes a 
second-class
constraint because of the boundary effect; because of this fact, the
Dirac bracket can be explicitly
constructed without introducing additional gauge conditions contrast
to the boundary-less case. Following the Noether procedure,
the conserved charge is calculated, which contains the surface integral
term ($Q_S$) as well as bulk term ($Q_B$). Functional
variations of $Q_B$ and $Q_S$ have the boundary contributions but
their sum $Q (=Q_B +Q_S$) has no boundary contributions. However,
$Q_B$ and $Q_S$ as 
well as $Q$ are still differentiable contrast to recent claims of
Ba\~nados {\it et al.} [10, 16]. It is shown that the Poisson and Dirac bracket
algebras of the Noether charges $Q$ for the both gauge symmetry and {\it Diff}
symmetry ${\it accross}$ the boundary are the same and become the Kac-Moody 
(Sec. II) and Virasoro (Sec. III) algebras 
with classical central terms, respectively. 
In Sec. II. {\bf C}, the origin of the central terms is re-examined
and it is found that the unusual delta-function formulas, which contain the
full information of the boundary, are the essential source of the
(classical) center. Furthermore, in Sec. III. {\bf B} it is emphasized
that the CS theory provides an concrete realization of the {\it holography 
principle} and a
correspondence of the {\it three-dimensional CS theory with boundary/
one-dimensional conformal field theory} ($CS_{2+1}/CFT_1$), which is more
fundamental and generalizes the  conjectured correspondence of the 
three-dimensional anti-de Sitter space/two-dimensional conformal field theory
($AdS_{2+1}/CFT_2$). In Sec. IV, the 
equivalence to the symplectic reduction method is shown directly by
projecting the Dirac bracket of base fields onto the boundary. The 
connection to Ba\~nados's work is clarified also.
In Sec. V, the Yang-Mills-Chern-Simons (YMCS) theory is considered
as a generalization. Even with the sharp differences in the symplectic
structures and the Noether charges, the Kac-Moody algebra is
exactly the same as that of the CS theory. There is no Virasoro
algebra in this model because the {\it Diff} symmetry is broken by the
three-dimensional Yang-Mills term explicitly. In Sec. VI, a summary and several
applications and generalizations are discussed. In Appendix {\bf A},
it is shown that the 
(smearing) Gauss law constraint, which becomes a second-class
constraint when there is the 
boundary, satisfies the consistency condition of the Dirac's Hamiltonian
algorithm for both the CS and YMCS theories. In Appendix {\bf B}, the
symmetry algebras for the
{\it Diff} {\it along} the boundary, which has only the quantum
theoretical center, is 
considered. 

\begin{center}
{\bf II. Kac-Moody algebra of gauge transformation}
\end{center}

\begin{center}
{\bf A. Noether charge}
\end{center}

I start with the Chern-Simons Lagrangian on a two-dimensional disc $D_2$
\begin{eqnarray}
L_{CS}=\kappa \int _{D_2} d ^2x \epsilon ^{\mu \nu \rho} \left<
  A_{\mu} \partial 
_{\nu} A_{\rho} +\frac{2}{3} A_{\mu} A_{\nu} A_{\rho} \right >,
\end{eqnarray}
where $\left<\cdots\right>$ denotes trace.
Up to a boundary term, (1) 
can be put into the canonical form with the
Lagrangian
\begin{equation}
L_{CS}= \frac{1}{2} \kappa \int_{D_2} d^2 x \epsilon^{ij} ( - A_i^a
\dot{A}^{a}_{j} 
+ A_0^a F_{ij}^a).
\label{eq:2}
\end{equation}
(Here, $\epsilon^{012} \equiv \epsilon^{12}\equiv 
1,~ A_i =A^a_i t_a,~ F_{ij}=F_{ij}^a t_a,~F_{ij}^a=\partial_i A^a_j 
-\partial_j A^a _i +f^{abc} A_i^b A_j^c$, and the group generators $t^a$ 
satisfy $[t^a, t^b]=f^{abc} t^c,~\left<t^a t^b\right>=\frac{1}{2}
\delta^{ab}$.)
I shall take (\ref{eq:2}) as my starting point [2, 10, 12, 16].\footnote{The
  non-covariant Lagrangian form (2) is the simplest action which is
  gauge invariant and retains the usual (bulk) equations of motion (4):
  If the covariant Lagrangian is considered as the starting point, one
  must introduce the additional function $c(\varphi , t)$ which relates
  $A_0|_{\partial D_2} =c(\varphi , t) A_{\varphi} |_{\partial D_2}$
  [see Ref. [20] for comparison] where the variation with
  respect to $c(\varphi , t)$ produces the boundary condition (6). But,
  from the gauge invariance requirement, which is necessary in the
  discussions of the Kac-Moody algebra, additional condition
  `$A_0|_{\partial D_2}= \varphi$ independent' should be introduced in
  order not to obtain the trivial result of {\it zero center}.} 
Variation of the
Lagrangian (2) gives\footnote{Here, I define $A_r=\hat{r}^i
  A_i,~A_{\varphi}/r=\hat{\varphi}^i A_i$ for the radial, (polar)
  angular coordinate $r, \varphi$ and their corresponding orthogonal unit
  vectors $\hat{r},~\hat{\varphi}$ on $\partial D_2$.}
\begin{eqnarray}
\delta L_{CS} = \kappa \int_{D_2} d^2x \left< \delta A_{\rho}
    \epsilon^{\rho \mu 
    \nu} F_{\mu \nu} \right> +2 \kappa \oint_{\partial D_2} d \varphi
\left< A_0 \delta A_{\varphi} \right>.
\end{eqnarray} 
[ I neglect the total time derivative term which is removed by
considering action variation $\delta I=\int^{t_2}_{t^1} dt \delta L_{CS}$
and the usual boundary conditions $\delta A_i |_{t^1}=\delta A_{i}
|_{t^2}=0$. However, the total space derivatives term, which becomes the
boundary Lagrangian in (3), can not be removed by choosing $\delta
A_{\varphi} |_{\partial D_2} =0$: This boundary condition kills all
the local boundary degrees of freedom which is a dangerous situation.]
In order to get the usual equation of motion\footnote{Depending on the
  supplemented boundary Lagrangians and boundary conditions, the
  equations of motion can be modified by the boundary term [21]. 
But the consistency and equivalence to the theory with (4) is not clear.}
\begin{eqnarray}
F_{\mu \nu} =0,
\end{eqnarray} 
even when there is the boundary, I choose the boundary conditions [16]
\begin{eqnarray}
&&A_0|_{\partial D_2} \propto A_{\varphi}|_{\partial D_2}, \\
&&\oint _{\partial D_2} d \varphi \left< A_{\varphi} A_{\varphi} \right>
=\mbox{fixed}  
\end{eqnarray}
for each time $t$; actually this boundary conditions have no role in the
Kac-Moody algebra for the gauge transformation but important role in
the Virasoro algebra for {\it Diff} [11]. The spatial part of the
equations of motion (4) gives
the Gauss law constraint
\begin{eqnarray}
G^a =\frac{1}{2} \kappa \epsilon ^{ij} F_{ij}^a =0  
\end{eqnarray}
from the variation with respect to $A^a_0$, independently on the
boundary conditions (5) and (6). In the symplectic reduction method, the
analysis of symmetry algebras is carried out after the constraint (7)
is explicitly solved [11-13]. But I am
considering an alternative approach where the constraint is not solved
but imposed only after all the processes of analysis are completed 
$ a'~ la$ Dirac. It is widely believed that these two methods are 
equivalent, of course when they both can be applied, although there is
no {\it general} proof. However, the Dirac method will be
unique one when the constraint can not be solved like as in the YMCS
model which will be treated in Section V. Moreover, in the boundary
theory the equivalence is not 
trivial matter as will be shown in this paper which involves some
non-trivial facts. 

Now, I consider the time-independent gauge transformation which is generated by
\begin{eqnarray}
\delta A^a_i &=&D_i \lambda^a, \nonumber \\
\delta A^a_0 &=&f^{abc}A_0^b \lambda^c.  
\end{eqnarray}
($D_i$ is the covariant derivative $D^{ab}_i =\delta^{ab} \partial_i
+f^{abc}A_i^c.$) Under this transformation, the Lagrangian (2)
transforms as\footnote{Under the `large' (time-independent) gauge
  transformation, the Lagrangian (2) transforms as $L_{CS}
  \rightarrow L_{CS} +\frac{d}{dt} X$ with $X=\kappa \int _{D_2} d^2 x
  \epsilon^{ij} \left< \partial _i U U^{-1} A_j \right> $ and the
  quantization of $\kappa$ is not needed to recover the gauge
invariance even at the quantum level due to the trivial homotopy
group. However, for more general 
time-{\it dependent} gauge transformation, the Wess-Zumino term must
be introduced to attain the gauge invariance [2, 21]. The quantization of
$\kappa$ in this case remains unclear.}
\begin{eqnarray}
\delta L_{CS}=-\kappa \frac{d}{dt} \int _{D_2} d^2x \epsilon^{ij}
  \left< \partial_i 
  \lambda A_j \right> \equiv \frac{d}{dt} X.  
\end{eqnarray}
Then, the Noether charge associated with this gauge transformation is
given by 
\begin{eqnarray}
Q(\lambda) &=&-\frac{\delta L_{CS}}{\delta \dot{A}^a_j} \delta A^a_j +X
  \nonumber \\
           &=&\kappa \int_{D_2} d^2 x \epsilon^{ij} \left<F_{ij}
  \lambda \right> 
  -2 \kappa \oint_{\partial D_2} d \varphi \left< A_{\varphi} \lambda
  \right> \nonumber \\ 
           &\equiv& Q_B (\lambda) +Q_S (\lambda), 
\end{eqnarray}
where $Q_B (\lambda)$ and $Q_S (\lambda)$ are the bulk and surface
integration terms, 
respectively.

\begin{center}
  {\bf B. Poisson and Dirac bracket algebras of Noether charge}
\end{center}

The basic Poisson bracket which can be directly read off from the
symplectic structure of the Lagrangian (2) is [11]
\begin{eqnarray}
  \{ A^a_i(x), A^b_j(y) \} &=&\frac{1}{\kappa} \epsilon^{ij}
 \delta^{ab} \delta^2 (x-y), \\
 \mbox{others} &=&0, \nonumber
\end{eqnarray}
and the Poisson bracket of any two function(al) $A, B$ is given by
\begin{eqnarray}
  \{ A, B \} =\int_{D_2}d^2 z \frac{\delta A}{\delta A^a_i (z)}
  \frac{\epsilon^{ij}}{\kappa} \frac{\delta B}{\delta A^a_j(z)}.
\end{eqnarray}
For the Noether charge $Q$ and its two constituents $Q_B$ and $Q_S$,
the functional derivatives are calculated 
simply by considering the functional variations for the field $A^a_i$:
\begin{eqnarray}
  \label{eq:13}
  \delta Q &=& \frac{1}{2} \kappa \int _{D_2} d^2 x \epsilon ^{ij}
  \delta F^a_{ij} 
  \lambda^a -\kappa \oint _{\partial D_{2}} d \varphi \delta
  A^a_{\varphi} \lambda^a 
  \nonumber \\
           &=&\kappa \int _{D_2} d^2 x \epsilon^{ij} \delta A^a_i (D_j
  \lambda)^a, \\  
  \delta Q_B &=& \kappa \int _{D_2} d^ x \epsilon^{ij} \delta A^a_i
  (D_j \lambda)^a + 
             \kappa \oint _{\partial D_{2}} d \varphi \delta
  A^a_{\varphi} \lambda^a 
  \nonumber \\
            &=&\kappa \int _{D_2} d^2 x \left[ \epsilon^{ij} \delta
  A^a_i (D_j \lambda)^a  
             + \delta(r-a) \delta A^a_i \hat{\varphi}^i \lambda^a
  \right], \\
 \delta Q_S &=& -\kappa \oint _{\partial D_2} d \varphi \delta
  A^a_{\varphi} \lambda^a 
  \nonumber \\
            &=&- \kappa \int _{D_2} d^2 x \delta(r-a) \delta A^a_i
  \hat{\varphi}^i \lambda^a. 
\end{eqnarray}
I note that $Q_B$ and $Q_S$ as well as their sum $Q$ have the well-defined
functional variations contrast to recent claims [10, 16]. ( `$a$' is
the radius of the boundary circle ${\partial D_2}$) Then, the
functional derivatives become 
\begin{eqnarray}
 \frac{\delta Q}{\delta A^a _i} &=& \kappa \epsilon^{ij} (D_j \lambda
 )^a ,\nonumber 
 \\
\frac{\delta Q_B}{\delta A^a _i} &=& \kappa \epsilon^{ij} (D_j \lambda )^a 
 +\kappa \delta(r-a) \hat{\varphi}^i \lambda^a , \\
\frac{\delta Q_S}{\delta A^a _i} &=& -\kappa \delta(r-a) \hat{\varphi}^i
 \lambda^a. \nonumber 
\end{eqnarray}
One notes that the derivatives of $Q_B$ and $Q_S$ have the boundary
effect terms which appear only for the variation on the
boundary. Then, using the formula (12) and the result (16),
it is easy to show the Poisson algebras of the $Q$'s as follows
\begin{eqnarray}
  \label{eq:17}
  \{Q_B(\lambda), Q_B(\eta) \} &=& Q_B ([\lambda, \eta] )-2 \kappa
  \oint _{\partial D_2} d \varphi\left< \lambda
  D_{\varphi} \eta \right> , \\
  \{Q_S(\lambda), Q_S(\eta) \} &=&0 ,\nonumber \\
  \{Q_B(\lambda), Q_S(\eta) \}&=&\{Q_S(\lambda), Q_B(\eta) \} \nonumber
  \\
                              &=&2 \kappa \oint_{\partial D_2} d
  \varphi \left< \lambda D_{\varphi} \eta 
  \right>, \nonumber \\
   \{Q(\lambda), Q(\eta) \} &=&Q([\lambda, \eta]) +2 \kappa
  \oint_{\partial D_2} d \varphi \left<
  \lambda \partial _{\varphi} \eta \right>,
\end{eqnarray}
where $[\lambda, \eta ]^a =f^{abc} \lambda^b \eta^c $. This results
show the Kac-Moody algebra with the central term for the
charge $Q$ even at the Poisson bracket level; however, I note that,
the Poisson algebra of $Q_B$ as well as $Q_S$ is not the Kac-Moody
algebra. In general, the algebras
will be modified by considering the Dirac bracket but except one case which
has been studied in Refs. [22]; as will be shown later this is actually
the exceptional case but now let me first consider the Dirac bracket algebra
explicitly following the usual Dirac's procedures.

To this end, it is important to note that the bulk charge
$Q_B (\lambda)$, which is a smearing quantity of the Gauss law constraint
$G^a = 0$ of (7) with the smearing function $\lambda$, becomes a
second-class constraint for the function $\lambda$ whose (angular)
derivative $\partial_{\varphi} \lambda$ as well as $\lambda$ itself  does not
vanish on the boundary [$2 \kappa \oint_{\partial D_2} \left< 
     \lambda D_{\varphi} \eta \right>=Q_S ([\lambda, \eta] )+ 2 \kappa
   \oint_{\partial D_2} \left<\lambda \partial_{\varphi} \eta \right>$
   ]; when $\lambda, \partial _{\varphi} \lambda $ vanish on the
   boundary, the theory becomes a trivial bulk one and this situation
   is not what I want to study. Now, it is found that the second-class
   constraint algebra comes from only 
the boundary effect and so additional gauge conditions are
not needed contrast to recent claims [10, 16]\footnote{This fact is
  related to that of the non-degeneracy of the symplectic structure of
  the boundary Lagrangian which is reduced from the original
  Lagrangian (2) by imposing the Gauss law constraint (7)}. (See the
Appendix {\bf 
  A} about the consistency with the Dirac's 
algorithm, {\it i.e.}, $\{Q_B, H_c\} \cong 0$ without introducing 
additional (secondary) constraints.\footnote{I thank Prof. R. Jackiw who first
asked about this problem.}) Then, the Dirac bracket of any two 
function(al) $A,B$ is defined by\footnote{I thank Prof. P. Oh who motivated for
me to consider the Dirac bracket explicitly}
\begin{eqnarray}
 \{ A, B \}^* =\{ A, B \} -\int [du][dv] \{ A, Q_B(u) \}
 \Delta^{-1}(u,v) \{ Q_B(v), B\}, 
\end{eqnarray}
where $\Delta^{-1}$ is defined as the functional inverse of
$\Delta(\lambda ,
\eta) \equiv \{ Q_B (\lambda), Q_B (\eta) \} \cong -2 \kappa
\oint_{\partial D_2} d \varphi \left<
  \lambda D_{\varphi} \eta \right>$ which depends, eventually, only on
the functions $\lambda, \eta, A_{\varphi}$ which live only on the
boundary: $\int [du] \Delta (\lambda, u) 
\Delta^{-1}(u, \eta) =\int [du] \Delta^{-1}(\eta, u) \Delta (u, \lambda)
=\delta (\lambda -\eta)$\footnote{Here, it would be more correct to
  confine $\lambda, \eta, ... etc.$ which live only on the boundary.
  However, since all the calculations
  involving $\Delta, \Delta ^{-1}$ are performed on the boundary
  $\partial D_2$, this rather formal definition also works as well. I thank
  Prof. S. Carlip for discussion about this matter}. [Weak equality
`$\cong$' means the equality after 
implementation of the constraint $Q_B=0$.] This bracket satisfies
\begin{eqnarray}
  \label{eq:20}
  \{ Q_B, B \}^* \cong 0
\end{eqnarray}
for any function(al) $B$ and so the Gauss law constraint $Q_B=0$ can be imposed
consistently in the Hamiltonian dynamics. (Here, it is difficult to
find the explicit solution of 
$\Delta^{-1}$ although it can be argued that this actually
exists\footnote{The matrix $\Delta(u,v)$, which is defined in the
  space of $u,v$, which live only on the boundary more correctly, has
  the non-vanishing determinant $\mbox{det} \Delta 
  (u,v) =(\Delta(u,v))^2 \neq 0$ ( $u,v$  are treated as the indices
  of the matrix and $\Delta(u,u)=0$ is used) unless one considers a
  trivial (bulk) theory of $\Delta(u,v)=0$.}
. But
I don't need the explicit solution in the main issue of this paper.) With this
Dirac bracket, it is easy to calculate the charge algebras as follows
\begin{eqnarray}
  \{Q_S(\lambda), Q_S(\eta) \}^* &\cong&Q_S([\lambda, \eta]) +2 \kappa
  \oint_{\partial D_2} d \varphi \left<
  \lambda \partial _{\varphi} \eta \right> , \\
\{Q(\lambda), Q(\eta) \}^* &\cong&Q([\lambda, \eta]) +2 \kappa
  \oint_{\partial D_2} d \varphi \left<
  \lambda \partial _{\varphi} \eta \right>.  
\end{eqnarray}
Here, one can observe the Dirac bracket algebra (22) of $Q$ is the same as the
corresponding Poisson algebra (18) but not for others  $Q_B$ and
$Q_S$: (21) can be
considered as the result of implementation of $Q_B=0$ in (22) but it
is different from the corresponding Poisson algebra (17). This peculiar
property of $Q$ can be also found in its generating gauge transformation:
The gauge transformation generated by the charges in the Poisson brackets
are given by
\begin{eqnarray}
 \{Q_B(\lambda), A^{a i}(x) \} &=&(D^i \lambda)^a +\xi^{a i}(\lambda),
 \nonumber \\
 \{Q_S(\lambda), A^{a i}(x)&=&-\xi^{a i}(\lambda), \\
\{Q(\lambda), A^{a i}(x) \} &=&(D^i \lambda)^a,
 \nonumber 
\end{eqnarray}
where $\xi^{a i}(\lambda)=\epsilon_{ij} \hat{\varphi}^j \delta(|{\bf x}|-a)
\lambda^a$ is the gauge transformation {\it only on the
boundary} \footnote{Temporal gauge transformation in (8) can be also
obtained by including $-\frac{\delta L_{CS}}{\delta A^a_0} \delta A^a_0
=-\int \pi^a_0 f^{abc} A^b_0 \lambda^c$ to the formula of the Noether
charge (10). But, since this additional
term is not important in my discussion, I will not consider this in
this paper.}: $Q_S$ generates only the {\it boundary} gauge transformation
$-\xi^{ai}$ which cancels that of $Q_B$, and $Q=Q_B+Q_S$ generates the
usual {\it bulk} gauge transformation even with boundary. The
corresponding ones in the Dirac bracket are
\begin{eqnarray}
\{Q_B(\lambda), A^{a i}(x) \}^* &\cong& 0, \nonumber \\
\{Q_S(\lambda), A^{a i}(x) \}^* &\cong&\{Q(\lambda), A^{a i}(x) \}^*
\cong(D^i \lambda)^a.   
\end{eqnarray}
So, one can find again the algebras involving $Q$ are the same
for the Poisson and Dirac brackets: But here, the bulk charge $Q_B$ is frozen
 and does not generate the gauge transformation;
instead, surface charge $Q_S$ acts like as the true generator of the full bulk
gauge transformation (8). This result means that the full bulk gauge
degrees of freedom for the system without boundary are transferred
completely into the boundary gauge degrees of freedom: Hence, this
CS theory with boundary can be considered as
a concrete realization of the {\it
  holography principle} which states `bulk world is an image of data
that can be stored on a boundary projection' [23]. Because of the
connection of the CS theory to the diverse areas of physics, the
principle can be 
applied more widely than currently limited cases of anti-de Sitter space
[24], like as in (Kerr-) de
Sitter space [17].

I conclude this subsection by summarizing that both the Poisson algebra of $Q$
and Dirac algebra of $Q$ (or $Q_S$) show the Kac-Moody algebra with
classical central term $2 \kappa \oint_{\partial D_2} \left< \lambda \partial 
_{\varphi}\eta \right> =\kappa \oint_{\partial D_2} \lambda^a
\partial_{\varphi} \eta^a$ and noting
that the existence of the central term is the purely Abelian effect
with no mixing between different colors.

\begin{center}
{\bf C. Re-examining the origin of the central terms}  
\end{center}

Up to now, the calculation has been straightforward and the appearance of
the central term seems not to be so strange unless $\lambda$'s and their
derivatives $\partial_{\varphi} \lambda$ 
vanish on the boundary. But as will be shown in this Section these
conditions imply the unusual formulas 
of delta-function which can not seen in the calculation of
Section {\bf B}: In Section {\bf B}, one has observed only some
remnant effects of these unusual formulas which contain the full
informations about the boundary. Here, I only consider the Abelian case
for simplicity because the non-Abelian properties have no important
role.  To this end, I first note that the Poisson algebra of $Q_B$'s,
without using the formulas 
(13)-(16) but only the basic Poisson bracket (11), becomes
\begin{eqnarray}
  \label{eq:25}
  \{Q_B(\lambda), Q_B(\eta) \}&=&\frac{1}{4}\kappa \int_{D_2} d^2x \int
  _{D_2'} d^2 x'
  \epsilon^{ij} \epsilon^{kl} \{F_{ij}(x), F_{kl}(x') \} \lambda(x)
  \eta(x') \nonumber \\
      &=&\kappa \int_{D_2} d^2 x \int_{D_2'} d^2 x'  \epsilon^{ij}
  \partial_i \partial_j ' 
  \delta^2(x-x') \lambda(x) \eta(x').
\end{eqnarray}
Here, if one uses the usual formula for the derivative of delta-function
\begin{eqnarray}
  \label{eq:26}
  \partial'_j \delta^2(x-x') =-\partial_j \delta^2 (x-x'),
\end{eqnarray}
(25) will vanish trivially. But actually this formula (26) is not true
in this case 
and rather this depends on the smearing functions $\lambda, \eta$
which are the test functions of the $\partial_i \delta ^2 (x -x') $ in
(25); for example, the radial part of 
(26) is modified as
\begin{eqnarray}
  \label{eq:27}
  \hat{r}'^i \partial'_i \delta^2 (x-x') =- \hat{r}^i \partial_i
  \delta^2 (x-x') +\delta^2 (x-x') \delta (r-a)
\end{eqnarray}
or in an integral form with the smearing (test) function $\eta$
\begin{eqnarray}
 \int_{D_2} d^2 x' \hat{r}'^i \partial'_i \delta^2 (x-x') \eta(x') =-
 \hat{r}^i \partial_i \eta(x) +\delta (r-a)\eta(a, \varphi) \nonumber,
\end{eqnarray}
by carefully treating the boundary terms in the process of integration
by parts for the smearing function $\eta$ which does not vanish
on the boundary $r=a$. The angular part is not modified if $\eta$ is
single-valued function $\eta (r, \varphi=2 \pi)=\eta(r, \varphi=0):
\hat{\varphi}'^i \partial'_j \delta^2 (x-x')=-
\hat{\varphi}^i  \partial_j \delta^2 (x-x')$.

Moreover, the quantity $\epsilon^{ij} \partial_i \partial_j ' \delta^2
(x- x')$ does not vanish if its test function $\eta$ is not constant
on the boundary: In 
an integral form it becomes
\begin{eqnarray}
\int _{D_2} d^2 x' [ \epsilon^{ij} \partial_i \partial_j ' \delta^2
(x- x') ] \eta (x') =- \delta  (r-a)\hat{\varphi}^i \partial_i 
\eta( a, \varphi).
\end{eqnarray}
Now then, with formula (28) or by carefully treating the boundary
terms in the process of integration by part,
one can find that
\begin{eqnarray}
 \{Q_B(\lambda), Q_B(\eta) \} &=& \kappa\int_{D_2} d^2 x \lambda(x) \partial _i
 \int_{D_2 '} d^2 x'
\epsilon^{ij} \partial'_j \delta^2(x-x') \eta(x') \nonumber \\
                             &=&-\kappa \oint_{\partial D_2} d \varphi \lambda
                             \partial_{\varphi} \eta
\end{eqnarray}
which is an Abelian result of (17). So, one can find that the
existence of classical central term implies that the usual
delta-function formulas need the boundary 
corrections\footnote{The discrepancy between
  the usual formula (26) and the integration by parts of (25) was
  observed by Balachandran {\it et al.} [8] but he did not provide the
  complete solution. For a related early work about the formula in
  other contexts, see Ref. [25]; I thank Dr. K. Bering for informing
  this reference.}; actually the boundary terms in (13)-(16) are
the 
ramnent effects of these corrections. All the other algebras (17)-(24) can be
calculated in this way but after more tedious calculations than
previous calculations of Sec. {\bf B}.

\begin{center}
{\bf III. Virasoro algebra of diffeomorphism } 
\end{center}
\begin{center}
  {\bf A. Noether charge }
\end{center}

The CS Lagrangian (2) has not only the gauge symmetry but also
the other class of symmetry, {\it Diff} symmetry which involves the
reparametrization of the geometrical coordinates. As is well-known
[13, 26] {\it Diff} symmetry is an expected one because this corresponds to a
{\it field-dependent} gauge transformation when the equations of motion of
the (pure) CS Lagrangian, $F_{\mu \nu}^a =0$ is used. But it is not
straightforward to obtain the {\it classical} central term for the
{\it Diff} symmetry algebra 
(Virasoro central term) from the central term for the gauge
transformation (Kac-Moody central term): In the derivation of the
Kac-Moody algebras (18), (21), (22) the existence of the central term does
not depend on what boundary conditions one chooses for $\lambda$'s
only if the $\lambda$'s are non-constant and single-valued
functions on the boundary. However, in the derivation of the Virasoro
algebra, the existence of central term will depend crucially on the
boundary condition; in other words, the (classical) Virasoro algebra
can not be anticipated simply from the Kac-Moody algebra. There are
several possible 
boundary conditions which allow the {\it Diff} symmetry but since I am
only interested in the Virasoro 
algebra with the center, I will consider only one and unique boundary
condition which allows the Virasoro central term.
To this end, I start with the Lagrangian (2) and study the response of
$L_{CS}$ to a 
spatial and time-independent {\it Diff}:
\begin{eqnarray}
\delta_f x^{\mu} &=&-\delta^{\mu}_{~ k} f^k, \nonumber \\
\delta_f A^a _i &=&f^k \partial_k A^a_i +
(\partial_i f^k) A_k^a, \nonumber \\
\delta_f A^a _0 &=&f^k \partial_k A^a_0.
\end{eqnarray}
Under (30), one finds
\begin{eqnarray}
\delta_f L_{CS} &=& \kappa \int_{D_2} d^2 x  \epsilon^{ij} 
\partial _k \left< -f^k A_i \dot{A}_j+f^k A_0 F_{ij}\right> \nonumber \\
&=& -\kappa \oint_{\partial D_2} 
d \varphi f^r \left<A_r \dot{A}_{\varphi}
 -\dot{A}_r A_{\varphi}- A_0 \epsilon^{ij} F_{ij}\right>.
\end{eqnarray}
Now, one has two possible boundary conditions in order that there are
{\it Diff} invariance, i.e.,  $\delta_f L_{CS}=\frac{d}{dt} X$:
(a). $f^r |_{\partial D_2}=0$, (b). $A^a_r |_{\partial D_2}$
=constant. [Remember that one needs the boundary conditions (5), (6)
already. Since the condition (6) is equivalent to $\partial_r
A^a_{\varphi}|_{\partial D_2} =0$, the condition (b) is exactly the same
as that of the symplectic reduction method [12].]. But, the condition
(a) does not produce the Virasoro algebra with center and I will concentrate
only the more interesting case (b) which produces the Virasoro center. (See
Appendix {\bf B} for analysis of case (a)) From
the condition (b), (31) becomes $\frac{dX}{dt}$ with 
$X=-\frac{1}{2} \kappa\oint_{\partial D_2} d \varphi 
f^r A^a_r A^a_{\varphi}$. 
The Noether charge becomes
\begin{eqnarray}
Q (f) &=-&\frac{\partial L_{CS}}{\partial \dot{A}^a_i} \delta_f A^a_i 
+X \nonumber \\
&=&\kappa \int_{D_2} 
d^2 x \left<f^k A_k \epsilon^{ij} F_{ij} \right> - \kappa \oint_{\partial
  D_2} d \varphi \left<2 f^r A_r A_{\varphi} 
+f^{\varphi} A_{\varphi}A_{\varphi} \right>  \\
  & \equiv & Q_{B} (f) + Q_{S} (f), \nonumber
\end{eqnarray}
where $Q_B (f)$ and $Q_S (f)$ are the bulk and surface
integration terms, respectively as in (10). [The Noether charges for
the {\it Diff} are distinguished from those of the gauge transformation
by the Roman parameters $f,g, ... etc$.]

\begin{center}
{\bf B. Dirac bracket algebra of Noether charge}  
\end{center}

By noting that there is no functional variation of ${A^a_r}|_{\partial
D_2}$ because of the boundary condition `$ {A^a_r}|_{\partial
D_2}$=constant'
in the last Section, the functional variations of the Noether charge
$Q$ and its bulk and surface constraints $Q_B, Q_S$ become
\begin{eqnarray}
\delta Q ({f}) &=& \kappa \int _{D_2}  d^2 x \left[ \epsilon^{ij}
  \delta A^a_i D_j (A_k 
  f^k)^a +\frac{1}{2} \epsilon^{ij} F^a_{ij}\delta A^a_{k} f^k \right],
  \\
\delta Q_{B} (f) &=&\kappa \int _{D_2} d^2 x  \left[ \epsilon^{ij}
  \delta A^a_i D_j (A_k 
  f^k)^a +\frac{1}{2} \epsilon^{ij} F^a_{ij}\delta A^a_{k} f^k \right]
+\kappa \oint _{\partial D_2} d \varphi \delta A^a_{\varphi} A^a_{k}
  f^k ,
  \\ 
\delta Q_{B} (f) &=&-\kappa \oint _{\partial D_2} d \varphi \delta
  A^a_{\varphi} A^a_{k} f^k , 
\end{eqnarray}
and their functional derivatives become
\begin{eqnarray}
\frac{\delta Q (f)}{\delta A^a_i} &=& \kappa \epsilon^{ij} \delta
  A^a_i D_j (A_k 
  f^k)^a +\frac{1}{2} \kappa \epsilon^{jk} F^a_{jk}f^i ,  \nonumber
  \\
\frac{\delta Q_{B} (f)}{\delta A^a_i} &=& \kappa \epsilon^{ij} \delta
  A^a_i D_j (A_k 
  f^k)^a +\frac{1}{2}\kappa \epsilon^{jk} F^a_{jk}f^i + \kappa \delta(r-a)
  \hat{\varphi}^i A^a_k f^k , 
  \\
\frac{\delta Q_S (f)}{\delta A^a_i} &=& -\kappa \delta(r-a)
  \hat{\varphi}^i A^a_k f^k . \nonumber
\end{eqnarray}
Here, I note that the functional variations (33)-(35) and derivatives
(36) for {\it Diff} is exactly the same form as those of gauge
transformations (13)-(15) and (16), respectively with the
field-dependent gauge transformations with the transformation function 
$\lambda^a =A^a_k f^k$ although the charges themselves are not exactly the
same  forms; this is an easily
expected result for $Q (f)$ since the {\it Diff} (30), which {\it will} be
generated by $Q (f)$, are easily expressed as a field
dependent gauge transformation up to the equations of motion term [26]
\begin{eqnarray}
  \delta_f A^a_i =D_i (f^k A_k)^a +f^k F_{ki}^a .
\end{eqnarray}

Using the formula (12) and the result (36), one finds the Poisson
algebras of the $Q ({f})$'s as follows
\begin{eqnarray}
\{ Q_{B} (f), Q_{B} (g) \} &=&-2 \kappa \oint _{\partial D_2} d
  \varphi \left< A_k f^k 
  D_{\varphi} (A_l g^l) \right> \nonumber \\
 &&+ 2 \kappa \int_{D_2} d^2 x \left< [ A_k, A_l] f^k
  g^l G + ({\bf f} \times {\bf g}) ~G \cdot G  
 -(A_k f^k D_j (G g^j) -(f \leftrightarrow g)) \right> \nonumber \\
 &\cong& -2 \kappa \oint _{\partial D_2} d \varphi \left< A_k f^k
  D_{\varphi} (A_l g^l) 
  \right> , \\
\{ Q_S (f), Q_{S} (g) \} &=& 0, \nonumber \\
\{ Q_{B} (f), Q_{S} (g) \} &=& \{ Q_S (f), Q_B (g) \}\nonumber \\
 &=& 2 \kappa \oint _{\partial D_2} d \varphi \left< A_k f^k
  D_{\varphi}(A_l g^l) - a 
  A_k g^k f^r G \right> \nonumber \\
 &\cong& 2 \kappa \oint _{\partial D_2} d \varphi \left< A_k f^k
  D_{\varphi} (A_l g^l) 
  \right> , \\
\{ Q (f), Q (g) \} &=&2 \kappa \oint _{\partial D_2} d \varphi \left< A_k f^k
  D_{\varphi} (A_l g^l) \right> \nonumber \\
 &&+ 2 \kappa \int_{D_2} d^2 x \left< [ A_k, A_l] f^k
  g^l G + ({\bf f} \times {\bf g}) ~G \cdot G - 
 (D_j (A_k f^k) G g^j -(f \leftrightarrow g)) \right> \nonumber \\
 &\cong& 2 \kappa \oint _{\partial D_2} d \varphi \left< A_k f^k
  D_{\varphi} (A_l g^l) 
  \right>. 
\end{eqnarray}
Using the fact [9, 10, 22] ($[f,g]^k_{Lie} =f^{n} \partial
_{n} g^k -g^{n} \partial_{n} f^k$ is Lie bracket\footnote{By
  considering the {\it Diff} of $A^a_r |_{\partial D_2} $(= constant)
  which reads 
  $0=\delta _f A^a_r |_{\partial D_2} =[D_r (f^k A_k)^a +f^k
  F_{kr}^a]|_{\partial D_2}=
  [(\partial _r f^r) A^a_r +(\partial_r f^{\varphi})
  A^a_{\varphi}]|_{\partial  D_2}$, one
  deduces an additional condition $\partial _r f^k |_{\partial
    D_2}=0$. From this fact, it is found that this Lie bracket is
  equivalent to $[f,g]^k_{Lie} =f^{\varphi} \partial
_{\varphi} g^k -g^{\varphi} \partial_{\varphi} f^k$ which is the Lie
bracket on the circle ($\partial D_2$) [12, 16]. This property will be
useful in the analysis of the higher-dimensional algebra, if there
is.}  
on the bulk ($D_2$) )
\begin{eqnarray}
\oint _{\partial D_2} d \varphi \left< A_k f^k \partial_{\varphi} (A_l
  g^l)\right>= 
\oint_{\partial D_2} d \varphi \left< [f,g]^{\varphi}_{Lie} A_{\varphi}
  A_{\varphi} + 2 [f,g]^r_{Lie} A_r A_{\varphi} +2 f^r
  \partial_{\varphi} g^r A_r A_r \right>
\end{eqnarray}
and $\left< A_k f^k D_{\varphi} (A_l g^l)\right>= \left< A_k f^k
  \partial_{\varphi} (A_l g^l)\right>$, the algebras become
\begin{eqnarray}
\{ Q_{B} (f), Q_B (g) \} &\cong& -Q_S ([f,g]_{Lie}) -2 \kappa \left< A_r A_r
  \right> \oint _{\partial D_2} d \varphi f^r \partial _{\varphi} g^r , \\
\{ Q_S (f), Q_S (g) \} &=& 0 , \nonumber \\
\{ Q_{B} (f), Q_{S} (g) \} &=& \{ Q_S (f), Q_{B} (g) \}\nonumber \\
  &\cong& Q_S ([f,g]_{Lie})+2 \kappa \left< A_r A_r
  \right> \oint _{\partial D_2} d \varphi f^r \partial _{\varphi} g^r , \\
\{ Q (f), Q (g) \} &\cong& Q_S ([f,g]_{Lie})+2 \kappa\left< A_r A_r
  \right> \oint _{\partial D_2} d \varphi f^r \partial _{\varphi} g^r. 
\end{eqnarray}
Here, there is no non-trivial non-Abelian effect which mixes the
different colors for the center: Similar to the
central term of (18), all the color degrees of freedom are simply
added up.

Now, using the Dirac bracket (19) the Dirac bracket algebra of $Q
(f)$'s, which has never been calculated explicitly due to the complications,
becomes\footnote{In the quantized theory, there is a further positive
  contribution to the center owing to the normal ordering
  effect for the unitary highest weight representation of Kac-Moody
  algebra with $\kappa > 0$ [10, 27]. Thus, $Q (f)$ does not produce the
  closed algebra, {\it i.e}., Wit algebra
  even when quantized. This situation is
  similar to the linear dilaton CFT and Liouville theory [28].}
\begin{eqnarray}
\{ Q (f), Q (g) \}^*\cong \{ Q_S (f), Q_S (g) \}^*
&\cong& Q_S ([f,g]_{Lie})+2 \kappa \left< A_r A_r
  \right> \oint _{\partial D_2} d \varphi f^r \partial _{\varphi} g^r.  
\end{eqnarray}
By considering a particular {\it Diff} with $f^r|_{\partial D_2}
\propto \partial_{\varphi} f^{\varphi}|_{\partial D_2}$, which is
required from the Jacobi identity [12], and the additional constant term
$\oint _{\partial D} d \varphi f^{\varphi} A_r^a A_r^a$, one finds that this
becomes the standard form of the Virasoro algebra after a proper normalization.
Similar to the gauge transformation of Sec. II, the Dirac bracket
algebra of $Q (f)$ in (45) is the same as the corresponding Poisson
algebra (44) although not the same for $Q_{B} (f)$ and $Q_S (f)$. The
{\it Diff} which is 
generated by the Noether charge of (34) becomes
\begin{eqnarray}
  \label{eq:47}
  \{ Q_{B} (f), A^{a i} \} &=&f^k \partial_k A^{a i} + (\partial ^i f^k
  )A^{a}_k + \xi^{a i}_f , \nonumber \\
\{ Q_S (f), A^{a i} \} &=&-\xi^{a i}_f ,\nonumber \\
\{ Q (f), A^{a i} \} &=&f^k \partial_k A^{a i} + (\partial ^i f^k
  )A^{a}_k  ,
\end{eqnarray}
where $\xi^{a i}_f =\epsilon_{ij} \hat{\varphi}^j \delta(|{\bf x}|-a)
A^a_k f^k$ is the {\it Diff} on the boundary. The corresponding
Dirac brackets are
\begin{eqnarray}
  \label{eq:47}
  \{ Q_{B} (f), A^{a i} \}^* &\cong& 0 ,\nonumber \\
\{ Q (f), A^{a i} \}^*&\cong& \{ Q_S (f), A^{a i} \}^* \cong f^k
\partial_k A^{a i} + (\partial ^i f^k 
  )A^{a}_k. 
\end{eqnarray}
Hence in the {\it Diff} case also, one finds the algebras involving
$Q (f) $ are the same for the Poisson and Dirac brackets and all the
information of {\it Diff} are stored in the surface charge $Q_S$ and the
bulk charge $Q_B$ is frozen in the Dirac bracket, coherently with the
holography principle. Moreover, one finds that the CS theory shows a
correspondence of the 
three-dimensional CS theory with boundary/one-dimensional conformal
field theory ($CS_{2+1}/CFT_1 $ ) which 
generalizes the conjectured $AdS_{2+1}/CFT_2$ correspondence [29]: The CS 
theory has one copy of the (real) Virasoro algebra and thus describes
a {\it one-dimensional} (real) conformal field theory. On the other hand, the
$AdS_{2+1}$, which can be constructed by the two copies of the $SL(2,
{\bf R})$ CS 
theories has two copies of the (real) Virasoro algebras and thus
describes two copies, {\it i.e.,} {\it two-dimensional} (real)
conformal field theory 
($CFT_{2}$). Hence, the $CS_{2+1}/CFT_1$ correspondence is more
fundamental than the conjectured $AdS_{2+1}/CFT_2$ correspondence; 
moreover the former correspondence derives also a correspondence of
the three-dimensional 
(Kerr-) de Sitter space/one-dimensional complex conformal field theory
(with the group $SL(2, {\bf C})$) [17] which has not been studied well.
 
\begin{center}
  {\bf IV. Relation to previous works}
\end{center}

There are two closely related works which used different methods to
obtain the anomalous symmetry algebras: One is the work done by Ba\~nados [10]
and the other is the work of Ref. [12].
In this Section, I discuss the relations between the calculation of
this paper and those of two previous works.

\begin{center}
  {\bf A. Relation to Ba\~nados's work}
\end{center}

Let me start first by recalling an interesting points which have been
emphasized in two previous Sections. It is the fact that the Poisson
bracket algebra of the Noether charge $Q=Q_B +Q_S$ itself is the same
as the corresponding Dirac bracket algebra for both the gauge
transformation and {\it Diff}. Actually, this 
is a peculiar situation in the Dirac method and it is known that there
is a unique case where this happens, in a recent analyses 
[22, 30]: Translated
into my case, the result says that
\begin{eqnarray}
\{ L_a, L_b \} \cong \{ L_a, L_b \}^*  
\end{eqnarray}
when $L_a$ is a quantity which commutes with the $Q_B$ in the Poisson
bracket, $\{ L_a, Q_B \}  \cong 0$ (which can be called
``gauge invariant quantity'' if there is no boundary [20]\footnote{ In
the previous case of Ref.[22, 30], $T (Q_B$ in this paper) was the 
first-class constraint and additional gauge fixing condition $\Gamma$ was
considered such that the formula (called master formula) becomes
$\{L_a, L_b\} \cong \{L_a, L_b \}^*_{\Gamma}$ which expressing the gauge
independence of the equality. However, the important thing is that the
validity of the formula is not limited to the first-class constraint
$T$.}). Now, since it is easy to observe that
\begin{eqnarray}
\{ Q, Q_B \} \cong 0 
\end{eqnarray}
for both the gauge and {\it Diff} transformations in the previous
Sections, it is now clear why 
the formula (48) works for $L \equiv Q$. In the literatures, a
formula which is essentially the same as (48) was assumed, implicitly
or explicitly, in
several models with an asymptotic boundary [4, 9]; the method which
has been used in these models was applied recently to the CS theory
with the finite and/or infinite boundary by Ba\~nados {\it et al.}
[10, 16] but it was unclear how the 
method could be applied to the the finite boundary, which has an
important meaning for understanding the nature of entropy of the
horizon space-times [2, 6, 16, 17], as well as to the infinite
boundary. Now,
it is clear why their method, which use (48) essentially, can be
applied to even the finite boundary systems
with the help of (49) and so the Ba\~nados's calculation method for the
boundary CS theory can be
justified. However, unfortunately I haven't been able to find any
general argument for the validity of (49) for the general Noether
charge $Q$ which produces the equality (48); if it is generally valid,
it will be a powerful tool for the evaluation of the Dirac bracket
when the straightforward calculation is difficult due to technical
reasons.  

On the other hand, the Noether charge $Q$ of (10) and (32) are the
same as the Ba\~nados's smeared generator\footnote{This construction
  was first considered by Regge-Teitelboim [9] and this has been used
  later, sometimes in their name [4, 9, 10].}  
\begin{eqnarray}
  \label{eq:50}
H(\eta)=\int _{D_2}  d^2 x \eta^a G^a +J(\eta),    
\end{eqnarray}
where the boundary term $J(\eta)$ is introduced (by hand) such that
$H(\eta)$ has no 
boundary terms in the functional variation\footnote{ This has been called
  `differentiability' but this will be miss-named one
  according to the diffentiablity even with the boundary terms in the
  functional variations.}. However, it is not evident
how the Noether charge and smeared generator are equivalent in
general: When $G^a$ represents the first-class constraints and one
restricts to the bulk symmetry transformation without the surface
transformation, the equivalence can be considered as a form of the
Dirac's conjecture [19], which states all the first-class constraints
(secondary as well as primary) become the symmetry generators,  which
has been widely believed without complete
proof\footnote{Recently, some proofs of the conjecture have been known
with several assumptions and explicit examples  
were also found, where the assumption were not valid [30, 31]}. However, my 
result implies that the equivalence may be valid even
for the second-class constraint $G^a$ although it is not found 
any formal proof for the validity similar to the case of (49).

\begin{center}
{\bf B. Relation to symplectic reduction method}
\end{center}

In the symplectic reduction method, the bulk Lagrangian (1) reduces to,
with the pure gauge solution $A_i=g^{-1} \partial _i g $,
\begin{eqnarray}
L_{CS}=-{\kappa} \int_{D_2} d^2 x \epsilon^{ij} \left< \partial_i 
g^{-1} \partial_j g g^{-1} \dot{g} \right> - {\kappa} 
\oint_{\partial
 {D_2}} d \varphi \left< g^{-1} \partial_{\varphi} g g^{-1} \dot{g} \right> 
\end{eqnarray}
which is essentially a boundary Lagrangian upon the local
parameterization of $g$ [1, 13], and its corresponding Poisson bracket
is
\begin{eqnarray}
\{A_{\varphi}^a(\varphi), A_{\varphi}^b (\varphi ') \}
&=&\frac{1} {\kappa}\left
(D_\varphi\delta(\varphi -\varphi^\prime)\right)^{ab}, \\
\mbox{others}&=&0 \nonumber
\end{eqnarray}
which are defined on the boundary $\partial D_2$. In this case, both
the base manifold and the symplectic structure of the Lagrangian are
drastically changed when the symplectic reduction is performed, but
the equivalence to the Dirac method was not well known\footnote{There
  is a known proof of the equivalence when the base manifold is not
  changed upon the symplectic reduction [30, 32]. The generalization of
  the proof to the changeable manifold will be interesting.}.
However, the results of the
previous Sections show the complete equivalence at least in the 
symmetry algebras [12] of the
boundary CS theory which can be considered as affirmative sign of the
equivalence even for the boundary theories. In this subsection I present
the more direct
equivalence proof in the basic bracket of fields $A^a_i$ from which the
equivalence of the charge algebras can be easily inferred. Following the
definition of (19), the Dirac bracket between $A^a_i$'s is given by
\begin{eqnarray}
\{A^a_i(x), A^b_j(x') \}^* =\frac{1}{\kappa} \epsilon^{ij} \delta^{ab}
\delta^2(x-x') + 
\int [du][dv] \left[ (D_i u)^a +\xi^a_i \right] \Delta^{-1}(u,v)
\left[ (D_j v)^b +\xi^b_j \right],
\end{eqnarray}
where the first equation of (23) is used; this is valid over all space
including boundary by construction. Now, let me project this bracket
onto the boundary $\partial D_2$
by multiplying $a \hat{\varphi}^i a \hat{\varphi}^j$:
\begin{eqnarray}
  \label{eq:52}
  \{A^a_{\varphi}(x), A^b_{\varphi} (x')\}^* &=&\{a A^a_i
  \hat{\varphi}^i (x), a 
  A^b_j \hat{\varphi}^j (x') \}^* \nonumber \\
      &=& \int [du] [dv] D_{\varphi} u^a \Delta^{-1}(u,v) D_{\varphi '} v^b
  \nonumber \\
      &=&\frac{1}{\kappa} (D_{\varphi} \delta(\varphi - \varphi ') )^{ab},
\end{eqnarray}
where I used the fact of $\xi_{\varphi} =a \hat{\varphi}^i \xi_i =0$. Moreover,
from `$A^a_r |_{\partial D_2} =\hat{r}^i A^a_i |_{\partial D_2}$=constant' the
bracket for $A^a_r$ vanishes on the boundary $\partial D_2$, {\it i.e.,}
\begin{eqnarray}
  \label{eq:53}
  \{A^a_{\varphi}(x), A^b_r(x') \}^* =\{A^a_{r}(x), A^b_r (x') \}^* =0.
\end{eqnarray}
These basic bracket algebras (54) and (55) are the same as (52) of the
symplectic reduction method [12] which solves the Gauss law from the
start and reduce the action (2) to the boundary action. Hence the
Dirac bracket of the all 
space (boundary as well as bulk) is reduced to the symplectically
reduced bracket on the boundary by projecting the Dirac bracket onto
the boundary. This proves the equivalence of the Dirac method and
symplectic reduction method at the fundamental level. This is the
first time to derive the boundary brackets (54), (55) directly from
the bulk bracket (11) as far as I know.
\begin{center}
{\bf V. Inclusion of Yang-Mills term }
\end{center}

So far, I have considered the pure CS term with the space boundary. In this
Section, I consider a generalized model with the Yang-Mills term
(Yang-Mills-Chern-Simons model (YMCS)) [7, 33, 34]. In this model, the
Dirac method is unique one because the equations of motion can not be
identically solved by the pure-gauge type solution contrast to the
pure CS theory such that the symplectic reduction method can not be applied.

I start with the YMCS Lagrangian on the disc $D_2$,
\begin{eqnarray}
  \label{eq:53}
  L_{YMCS} =\int_{D_2} d^2 x \left[-\frac{1}{4} F_{\mu \nu}^a F{^{\mu
  \nu}}^a -\frac{\kappa}{2} \epsilon_{ij} A^a _i \dot{A}^a _j +\frac{\kappa}{2}
  A_0 ^a \epsilon_{ij}F^a_{ij} \right],
\end{eqnarray}
where the CS part is the pure CS Lagrangian (2). Here, I note that the
added YM term break the {\it Diff} 
symmetry of the CS part although the gauge symmetry is preserved. So,
in this model, there is no Virasoro algebra but only the Kac-Moody
algebra [13, 26]. Furthermore, because of the YM term, the symplectic
structure of the total Lagrangian $L_{YMCS}$ is changed from the pure
CS Lagrangian: The basic
Poisson bracket is just the canonical one
\begin{eqnarray}
  \label{eq:54}
  \{A^a_i ( x), \pi^{bj} (y)\} &=&\delta_{ij} \delta^{ab} \delta^2 (x-y), \\
  \mbox{others}&=&0 , \nonumber
\end{eqnarray}
where $\pi^{ai} \equiv \frac{\delta L}{\delta \dot{A}^a_i}=F_{0i}^a
+\frac{\kappa}{2} \epsilon_{ij} A^a_j$ contrast to the pure CS case
(10): Although the Lagrangian $L_{YMCS}$ and $\pi^{ia}$ converge to
those of pure CS theory by $\kappa \rightarrow \infty$ limit, the
Poisson bracket does not; hence, it is not clear at the {\it
  algebraic} level whether the symmetry 
algebra of YMCS model converge into that of CS model in the large $\kappa$
limit or not.

Before considering the symmetry algebra, I first consider the variation
principle for the Lagrangian (56): The variation of the YM part of
(56) becomes (neglecting 
the total time derivative terms)
\begin{eqnarray}
  \label{eq:55}
  \delta L_{YM} &=&2 \int_{D_2} d^2 x \left< \delta A_{\rho} D_{\mu}
  F^{\mu \rho} \right> +2 \oint _{\partial D_2} d \varphi \left<
-\frac{1}{a} \left( \partial_r A_{\varphi}-\partial_{\varphi}A_r
  +[A_r, A_{\varphi}] \right) \delta A_{\varphi} \right. \nonumber \\
 &&+a\left. \left( \partial_r
  A_0 -\partial_0 A_r +[A_r, A_0 ] \right) \delta A_0 \right>.
\end{eqnarray}
So, by considering the total variation of $L_{YMCS}$ with $\delta
L_{CS}$ of (3), one can get the usual equations of motion
\begin{eqnarray}
  D_{\mu} F^{\mu \nu} +\frac{\kappa}{2} \epsilon^{\nu \mu \rho} F_{\mu
  \rho} =0
\end{eqnarray}
if one chooses the boundary conditions
\begin{eqnarray}
  &&A_0 |_{\partial D_2} =\mp \hat{\varphi} \cdot {\bf A} |_{\partial
  D_2} , \nonumber \\
  &&(\partial_0 \mp \hat{\varphi} \cdot \nabla ) A_r |_{\partial D_2} =0
\end{eqnarray}
as well as the condition (6). Here, I note that the 
condition (5) for the pure CS theory is more confined to the first
condition of (60), which corresponds to the horizon space-times in pure CS
gravity theory [16, 17] interestingly. The second condition represents
that $A_r$ is 
a transverse chiral mode along the boundary $\partial D_2$.

Now, returning to the symmetry algebra, it is noted that the YMCS
theory has the gauge 
symmetry with the same as the CS theory: Under the gauge
transformation (7), the YMCS Lagrangian transforms as $\delta L_{YMCS}
=\frac{d}{dt} X$ with $X=-\kappa \int d^2 x \epsilon_{ij}
\left<\partial_i \lambda A_j \right>$. Then, the associated Noether
charge is given by, according to (9),
\begin{eqnarray}
  Q (\lambda) &=&\int _{D_2} d^2 x \left< (2 D_j \pi^j +\kappa
  \epsilon_{ij} \partial_i A_j ) \lambda \right> -\oint _{\partial
  D_2} d \varphi \left<2 a \pi^r \lambda +\kappa A_\varphi \lambda
  \right> \nonumber \\
   &\equiv& Q_B (\lambda) +Q_S (\lambda),
\end{eqnarray}
where $Q_B (\lambda)$ and $Q_S (\lambda)$ are the bulk and surface
terms, respectively. 
Here, I note that $Q_B (\lambda) =2 \int_{D_2} d^2 x \left< \left( D_j F^{j0}
  +\frac{\kappa}{2} \epsilon^{0ij} F_{ij} \right) \lambda \right>$ which is a
smearing form of the Gauss law (0'th component) of (59). 

In parallel with the CS theory, the functional variations of the
Noether charge $Q$ and its constituents $Q_B, Q_S$ are calculated as
\begin{eqnarray}
  \delta Q &=&\int _{D_2} d^2 x \left[ \delta A^a_j \left(f^{abc} \pi^{jb}
  \lambda^a -\frac{\kappa}{2} \epsilon_{ij} \partial_i \lambda^a
  \right) -\delta \pi^{aj} \left( f^{abc} A^b_j \lambda^c +\partial_j
  \lambda^a \right) \right], \\
\delta Q_B &=&\int _{D_2} d^2 x \left[ \delta A^a_j \left(f^{abc} \pi^{jb}
  \lambda^a -\frac{\kappa}{2} \epsilon_{ij} \partial_i \lambda^a
  \right) -\delta \pi^{aj} \left( f^{abc} A^b_j \lambda^c +\partial_j
  \lambda^a \right) \right. \nonumber \\
       &+& \left. \delta(r-a) \left(\delta \pi^{a i} \hat{r}^i
  +\frac{\kappa}{2} \delta A^a_i \hat{\varphi}^i \right) \lambda^a
  \right], \\
 \delta Q_S &=&-\int_{D_2} d^2 x \delta(r-a) \left(\delta \pi^{a i} \hat{r}^i
  +\frac{\kappa}{2} \delta A^a_i \hat{\varphi}^i \right) \lambda^a . 
\end{eqnarray}
These all have the well-defined functional variations. Then, the functional
derivatives become
\begin{eqnarray}
\frac{\delta Q}{\delta A^a_{i}}& =&f^{abc} \pi^{b i} \lambda^c
       +\frac{\kappa}{2} \epsilon_{ij} \partial_j \lambda^a , \nonumber \\
\frac{\delta Q}{\delta \pi^{bi}} &=&-(D_i \lambda)^b , \nonumber \\
\frac{\delta Q_B}{\delta A^a_{i}}& =&f^{abc} \pi^{b i} \lambda^c
   +\frac{\kappa}{2} \epsilon_{ij} \partial_j \lambda^a +\delta(r-a)
   \frac{\kappa}{2} \hat{\varphi}^i \lambda^a , \nonumber \\  
\frac{\delta Q_B}{\delta \pi^{bi}} &=&-(D_i \lambda)^b
       +\delta(r-a)\hat{r}^i \lambda^b , \nonumber \\
\frac{\delta Q_S}{\delta A^a_{i}} &=&-\delta(r-a)\frac{\kappa}{2} 
\hat{\varphi}^i \lambda^b , \nonumber \\
\frac{\delta Q_S}{\delta \pi^{bi}} &=&-\delta(r-a) 
\hat{r}^i \lambda^b.
\end{eqnarray}
Using the result (65) and the Poisson bracket which is given by
\begin{eqnarray}
  \{ A, B \} =\int_{D_2} d^2 z \left( \frac{\delta A}{\delta A^a_i (z) } 
 \frac{\delta B}{\delta \pi^{ai} (z) }-
 \frac{\delta A}{\delta \pi^{ai} (z) } 
 \frac{\delta B}{\delta A^{a}_{i} (z) } \right),
\end{eqnarray}
the Poisson algebra of the $Q$'s becomes as follows
\begin{eqnarray}
  \label{eq:17}
  \{Q_B(\lambda), Q_B(\eta) \} &=&Q_B ([\lambda, \eta] )
    -Q_S ([\lambda, \eta] )-
   2 \kappa \oint _{\partial D_2} d \varphi\left< \lambda
  \partial_{\varphi} \eta \right>,  \\
  \{Q_S(\lambda), Q_S(\eta) \} &=&0 ,\nonumber \\
  \{Q_B(\lambda), Q_S(\eta) \}&=&\{Q_S(\lambda), Q_B(\eta) \} \nonumber\\
  &=&Q_S ([\lambda, \eta] )+ 2 \kappa \oint_{\partial D_2} d \varphi \left< 
     \lambda \partial_{\varphi} \eta \right>, \nonumber \\
   \{Q(\lambda), Q(\eta) \} &=&Q([\lambda, \eta]) +2 \kappa
   \oint_{\partial D_2} d \varphi
\left<\lambda \partial _{\varphi} \eta \right>.
\end{eqnarray}
This algebra is exactly the same {\it form} as
that of pure CS theory (17), (18) although the charges and the
basic symplectic structures are 
sharply different for the YMCS and pure CS theories.\footnote{This
result of $\{Q(\lambda), Q(\eta)
   \}$ agrees to the Dunne and Trugenberger's one [34]. But, they
   didn't find any good reason for including the surface term in
   $Q$. Mickelsson [33] only considered bulk part $Q_B$ and he found
   the correct central term for the first time but he missed the $Q_S$ term in
   (67).} 
Moreover, the large $\kappa$ limit is singular for the commutation
relation (65) although not for Q's themselves; on the other
 hand, the small $\kappa$ limit (pure Yang-Mills phase) is
 well-defined one which has vanishing center Virasoro algebra (Wit 
 algebra). The
 Dirac bracket is well-defined in the same as (19) from the
 second-class algebra of $Q_B$ also, and the Dirac 
 bracket algebra is the same as the pure CS theory: Moreover,
 since $\{Q_B(\lambda), Q(\eta) \} \cong 0$ also, the Poisson bracket
 algebra a nd
 Dirac bracket algebra are the same for the Noether charge Q's. The realization
 of holography principle is the same as that of CS theory due to the same
 gauge transformations (23) and (24). (See
 Appendix {\bf A}. (b) for the consistency with the Dirac algorithm.)

Before ending this Section, it seems to appropriate to note that the
angular projection of the 
Dirac bracket between $A^a_i$'s, $\{A^a_{\varphi}, A^b_{\varphi '}
\}^*$ is the same as that of pure CS theory. However, the Dirac
brackets 
containing $A^a_r$ can not be calculated completely without knowing
the explicit form
of $\Delta^{-1}$:
\begin{eqnarray}
  \{A^a_{\varphi} (x), A^b_{r'} (x') \}^* &=&D^{bc}_{r'} \int [du][dv]
  v^c D_{\varphi} u^a \Delta^{-1} (u,v) + \delta(r'-a) \int [du][dv]
  \Delta^{-1}(u,v) v^b (D_{\varphi} u)^a , \nonumber \\
\{A^a_{r} (x), A^b_{r'} (x') \}^* &=& \int [du][dv] \left[ (D_r u)^a
  -\delta(r-a) u^a \right] \Delta^{-1}(u,v)  \left[ (D_{r'} v)^a
  -\delta(r'-a) v^a \right]. 
\end{eqnarray}
This result implies that the symplectic structure involving
$A^a_{\varphi}$'s for the boundary YMCS theory is the same as the
boundary (pure) CS
theory, if one finds and considers the symplectic reduction which solves the
equations of motion (58), although not the same for other components.

\begin{center}
  {\bf VI. Summary and discussions}
\end{center}
It has been studied how the space boundary modifies
drastically the symmetry algebras in the pure CS theory and
YMCS theory. The most drastic one is the fact that the Gauss law constraint
$Q_B$, which was the first-class constraint in the usual theory where
the boundary effect was not considered, became the second-class
one; due to this fact, the
Dirac bracket was able to be constructed {\it explicitly}, which had
never been done 
previously from the lack of complete understanding of the constraints
structure, without introducing additional gauge conditions. Moreover,
the symmetry algebras of the Noether charges, Kac-Moody and 
Virasoro algebras, which had been known recently, with the
``classical'' centers have their origin in the non-commutability
of the Gauss law $Q_B$. In a mathematical terms, all these unusual things
were simple results of the unusual delta-function formulas which had
the boundary correction terms. Although it has been found that
the Dirac method is equivalent to the symplectic reduction method by
which the anomalous Ba\~nados algebra was explicitly derived first, the
previously noted peculiar properties were manifest only in the Dirac
method.

The boundary modified also the conserved Noether
charge by surface integral term $Q_S$ in addition to the usual bulk term
$Q_B$ and only the combination $Q_B +Q_S (=Q)$ generated the correct
(bulk) symmetry transformation. A peculiar result of this fact is
that the physical states of the quantum theory are not annihilated by the
symmetry generator $Q$ due to the non-vanishing part $Q_S$ which does
not constitute a constraint. Thus, the central term in the algebra of
$Q$ is not harmful in the quantization contrast to the usual (quantum)
anomaly of symmetry constraints. Moreover, it is important to note that
the second-class constraint algebra of $Q_B$ does not imply the
breaking of some symmetries: The only candidate of the broken symmetry
due to the boundary will be the time-dependent gauge (and {\it Diff}
also for the pure CS theory) symmetry by which the Lagrangian (2) and
(54) do not transforms as $\delta L =\frac{d}{dt} X$ and thus are not
gauge invariant. But, the center 
of the second-class algebra of $Q_B$ are independent on the
time-dependence of gauge transformation parameters $\lambda$ or $\eta$
and thus invalidate the connection of the non-commutability of $Q_B$ and gauge
non-invariance.

The holography principle {\it i.e.,} bulk theory/boundary theory
correspondence was also an interesting effect of the boundary. In this
paper, the $CS_{2+1}/CFT_1$ correspondence occurred and this is more
fundamental and can generalize the conjectured $AdS_{2+1}/CFT_2$
correspondence. 

Besides of these things, several things remains unclear. One is about
the explicit solution of $\Delta^{-1}$ although it was not needed in
many physically interesting cases\footnote{Similar situation is also
  occurred in the symplectic method. See the paper of Bak {\it el
    al.} [13].}. Second is about the question of the 
general validity of the property $\{Q, Q_B\} \cong 0~(49)$, which
makes the Poisson and Dirac bracket algebras for $Q$ be the same, for
the general Noether charge $Q$ and the bulk part $Q_B$. Third one is
the question about the $formal$ equivalence of the Noether procedure
and the Regge-Teitelboim procedure for constructing the symmetry
generators. Final one is the question about the origin for the same
symmetry (Kac-Moody) algebras of both CS and YMCS theories despite of
the sharp differences in the basic Poisson bracket and the Noether
charges.  

As the final remarks, it would be interesting to extend to
supersymmetric [35] and higher-dimensional CS theories [36] in
relation to the supergravity theories, higher-dimensional black hole
systems, M-theory and anticipated $CS_{d+1}/CFT_{d-1}$ correspondence:
Especially, in 
the higher-dimensional CS theories the use of Dirac method will be crucial in
the manipulation of the symmetry algebras because the general pure
gauge solution is not known in that case. Moreover, it is
interesting to find a transformation which transforms the 
different-dimensions CS  theories, which may reflect the $U$-duality of the 
D-brane configurations of the black holes [37].

\begin{center}
  {\bf Appendix A}
\end{center}
In this Appendix, I show that the second-class Gauss law (smearing)
constraint $Q_B \cong 0$ is consistent with the Dirac's Hamiltonian
algorithm, i.e., $\{ Q_B, H_c \} \cong 0$ without introducing 
additional (secondary) constraints for both pure CS and YMCS theories. 

{(a) pure CS theory}:

I start by noting that the canonical Hamiltonian of CS Lagrangian (2) becomes 
\begin{eqnarray}
  H_c &=& \frac{\kappa}{2} \int _{D_2} d^2 x  \left( -A_0^a \epsilon^{ij}
  F^a_{ij} \right) 
  \nonumber \\
      &=& Q_B (- A_0 ). 
\end{eqnarray}
Then, it is easy to show that
\begin{eqnarray}
  \{ Q_B, H_c \} &=&- Q_B ([\lambda, A^0 ]) -2 \kappa \oint _{\partial D_2} d
  \varphi \left< \lambda D _{\varphi} A_0 \right> \nonumber \\
             &=&- Q_B ([\lambda, A^0 ]) -2 \kappa \oint _{\partial D_2} d
  \varphi \left< \lambda \partial _{\varphi} A_0 \right> \\
                  &\cong& 0 \nonumber
\end{eqnarray}
if $\partial_{\varphi} A_0 |_{\partial D_2} \cong 0$ (restriction of
the Lagrange multiplier $A_0$) is satisfied. [In the second line, the
boundary condition (5) was used.] Hence, The
consistency condition [19] is satisfied without introduction of 
additional (secondary) constraints.

{(b) YMCS theory}:

In this case, the canonical Hamiltonian becomes

\begin{eqnarray}
H_c &=&\int _{D_2} d^2 x \left[ \frac{1}{2} E^{a i} E^{a i} +
  \frac{1}{2} B^a B^a - A^a_0 \left((D_i E^i)^a +\frac{\kappa}{2} \epsilon^{ij}
  F_{ij}^a \right) +\partial_i (A^a_0 E^{ai} ) \right] \nonumber \\
    &=& H_0 + Q_B(-A_0) +H_S , \\
 H_0 &\equiv& \int_{D_2} d^2 x \frac{1}{2} \left( E^{ai } E^{a i} +B^a
  B^a \right), \nonumber \\
 H_S &\equiv& \int _{D_2} d^2 x \partial_i (A^a_0 E^{a i}), \nonumber
\end{eqnarray}
where $B^a \equiv -\frac{1}{2} \epsilon^{ij} F_{ij}^a, E^{ai} \equiv
F^{0i a}$. Now, I need the functional derivatives of $H_0$ and $H_S$
in order to evaluate $\{Q_B, H_c \}$ in parallel with manipulation of
the text. The functional variations and their derivatives are as
follow:
\begin{eqnarray}
\delta H_0 &=&\int _{D_2} d^2 x \left[ E^{ai} (\delta \pi^{ai}
  -\frac{\kappa}{2} \epsilon^{ij} \delta A^a_j ) +(D_i B)^a
  \epsilon^{ij} \delta A_j^a -B^a \delta^A_i \hat{\varphi}^i \delta
  (r-a) \right] , \nonumber \\
\delta H_S &=&\int _{D_2} d^2 x \delta (r-a) \left[ \delta A^a _0 E^{ai}
  \hat{r}^ i + A^a_0 (\delta \pi ^{ai } -\frac{\kappa}{2} \epsilon^{ij}
  \delta A^a_j ) \hat{r}^i \right] , \nonumber \\
 \frac{\delta H_0}{\delta A^a_i}&=& \frac{\kappa}{2} \epsilon^{ij}
 E^{aj} -\epsilon^{ij}(D_j B)^a - B^a \hat{\varphi} ^i \delta (r-a) ,
 \nonumber \\
\frac{\delta H_0}{\delta {\pi}^{ai}}&=&  E^{ai} ,
 \nonumber \\
\frac{\delta H_S}{\delta A^a_i}&=& \delta (r-a) \frac{\kappa}{2}
\epsilon^{ij} \hat{r}^j A^a_0,
 \nonumber \\
\frac{\delta H_S}{\delta {\pi}^{a i}}&=& \delta (r-a) \hat{r}^i A^a_0.
 \nonumber \\
\end{eqnarray}
Then, one could show that
\begin{eqnarray}
\{ Q_B (\lambda), H_c \} = A + B + C , 
\end{eqnarray}
where
\begin{eqnarray}
 A&=&\{Q_B (\lambda), Q_B (-A_0) \} \nonumber \\
  &=&-Q_B ([\lambda, A_0 ]) +\oint _{\partial D_2} d \varphi \left< -2 a
  \pi^r [\lambda, A_0 ] +2 \kappa \lambda \partial _{\varphi} A^0
  \right> , \\
B&=& \int _{D_2} d^2 z \frac{\delta Q_B (\lambda)}{\delta A_i^a (z)}
  \frac{\delta (H_0 +H_S)}{\delta \pi^{ai} (z)} \nonumber \\
 &=&\int _{D_2} d^2 z \kappa \epsilon^{ij} \left< \lambda (D_i E^j)
  \right> + \oint _{\partial D_2} d \varphi \left< - \kappa \lambda
  \partial _{\varphi} A_0  +2 a E^r [\lambda, A_0 ] \right> , \\
C&=& -\int _{D_2} d^2 z \frac{\delta Q_B (\lambda)}{\delta {\pi}^{a i} (z)}
  \frac{\delta (H_0 +H_S)}{\delta A^{a}_{i} (z)} \nonumber \\
 &=&- \int _{D_2} d^2 z \kappa \epsilon^{ij} \left< \lambda (D_i E^j)
  \right> + \oint _{\partial D_2} d \varphi \left<  \kappa \lambda
  \partial _{\varphi} A_0  +2 \lambda (D_{\varphi} B) \right> .
\end{eqnarray}
Finally, one obtains
\begin{eqnarray}
\{ Q_B (\lambda), H_c \} &=& -Q_B ([\lambda, A_0 ])+ 2 \oint _{\partial
  D_2} d \varphi \left< \lambda ( \kappa 
  \partial _{\varphi} A_0  +D_{\varphi} B) \right> \\
           &\cong& 0 \nonumber 
\end{eqnarray}
if $\kappa \partial_{\varphi} A^0 +D_{\varphi} B =D_{\varphi} (\kappa
A^0 +B) \cong 0$ is satisfied. [Here, the boundary condition (5) was
used several times.] Hence, in the YMCS case also, the consistency
condition is satisfied without introduction of additional (secondary)
constraints. 

\begin{center}
  {\bf Appendix B}
\end{center}

Here, I present the Virasoro algebra for the {\it Diff} symmetry of the CS
theory with $f^r |_{\partial D_2} =0$ (i.e., {\it Diff} along the
boundary (${\partial D_2}$)) which does not produce the central term at
classical level. The Noether charge for this {\it Diff} becomes (X=0)
\begin{eqnarray}
Q (f) &=&\kappa \int_{D_2} 
d^2 x \left<f^k A_k \epsilon^{ij} F_{ij} \right> -\kappa \oint
_{\partial D_2} d \varphi \left<f^{\varphi} A_{\varphi}A_{\varphi}
\right>,  \\ 
  &\equiv& Q_{B} (f) + Q_S (f) \nonumber
\end{eqnarray}
with the bulk and surface terms $Q_B (f)$ and $Q_S (f)$
respectively. Actually, these charges $Q$'s can be directly obtained
by setting $f^r|_{\partial D_2}=0$ in (32). Moreover, all the other
algebras can be simply obtained by this reduction procedure from the
corresponding ones in Sec. III. {\bf B}: Since $f^r$ appears
always together with $A_r$ in the Noether charge and there is no
fundamental variation of
$A_r|_{\partial D_2}$ already, this reduction is well-defined. So, the
only modification is the surface term involving $f^r |_{\partial D_2}$
in the results of Sec. III. {\bf B}. The functional variation and
their functional derivatives of the Noether charge become
\begin{eqnarray}
\delta Q (f) &=&\kappa \int _{D_2}  d^2 x \left[ \epsilon^{ij} \delta
  A^a_i D_j (A_k 
  f^k)^a +\frac{1}{2} \epsilon^{ij} F^a_{ij}\delta A^a_{k} f^k \right] ,
  \\
\delta Q_{B} (f) &=&\kappa \int _{D_2} d^2 x  \left[ \epsilon^{ij}
  \delta A^a_i D_j (A_k 
  f^k)^a +\frac{1}{2} \epsilon^{ij} F^a_{ij}\delta A^a_{k} f^k \right]
+\kappa \oint _{\partial D_2} d \varphi \delta A^a_{\varphi}
  A^a_{\varphi} f^{\varphi} ,
  \\ 
\delta Q_{S} (f) &=&-\kappa \oint _{\partial D_2} d \varphi \delta
  A^a_{\varphi} A^a_{\varphi} f^{\varphi},  \\
\frac{\delta Q (f)}{\delta A^a_i} &=& \kappa \epsilon^{ij} \delta
  A^a_i D_j (A_k 
  f^k)^a +\kappa \frac{1}{2} \epsilon^{jk} F^a_{jk}f^i , 
  \\
\frac{\delta Q_{B} (f)}{\delta A^a_i} &=& \kappa \epsilon^{ij} \delta
  A^a_i D_j (A_k 
  f^k)^a +\kappa \frac{1}{2} \epsilon^{jk} F^a_{jk}f^i + \kappa\delta(r-a)
  \hat{\varphi}^i A^a_{\varphi} f^{\varphi},  
  \\
\frac{\delta Q_S (f)}{\delta A^a_i} &=& -\kappa \delta(r-a)
  \hat{\varphi}^i A^a_{\varphi} f^{\varphi}. 
\end{eqnarray}

The Poisson and Dirac bracket algebras of $Q (f)$'s are as follows
\begin{eqnarray}
\{ Q_{B} (f), Q_{B} (g) \} 
 &\cong& -2 \kappa \oint _{\partial D_2} d \varphi \left< A_{\varphi}
 f^{\varphi} D_{\varphi} (A_{\varphi} g^{\varphi}) 
  \right>, \\
\{ Q_S (f), Q_{S} (g) \} &=& 0 , \nonumber \\
\{ Q_{B} (f), Q_{S} (g) \} &=& \{ Q_S (f), Q_B (g) \}\nonumber \\
&\cong& 2 \kappa\oint _{\partial D_2} \left< A_{\varphi} f^{\varphi}
 D_{\varphi} (A_{\varphi} g^{\varphi}) 
  \right>, \\
\{ Q (f), Q ({g}) \} 
 &\cong& 2 \kappa \oint _{\partial D_2} \left< A_{\varphi} f^{\varphi}
 D_{\varphi} (A_{\varphi} g^{\varphi}) 
  \right> .
\end{eqnarray}
Now, using the fact
\begin{eqnarray}
\oint _{\partial D_2} \left< A_{\varphi} f^{\varphi} D_{\varphi} (A_{\varphi}
  g^{\varphi})\right>
&=&\oint _{\partial D_2} \left< A_{\varphi} f^{\varphi}
  \partial_{\varphi} 
(A_{\varphi} g^{\varphi})\right> \nonumber \\&=&
\oint_{\partial D_2} \left< [f,g]^{\varphi}_{Lie} A_{\varphi}
  A_{\varphi} \right>, \nonumber \\
\end{eqnarray}
the algebras become
\begin{eqnarray}
\{ Q_{B} (f), Q_B (g) \} &\cong& -Q_S ([f,g]_{Lie}) , \\
\{ Q_S (f), Q_{S} (g) \} &=& 0 , \nonumber \\
\{ Q_{B} (f), Q_{S} (g) \} &=& \{ Q_S (f), Q_B (g) , \}\nonumber \\
  &\cong& Q_S ([f,g]_{Lie}),  \\
\{ Q (f), Q (g) \} &\cong& Q_S ([f,g]_{Lie}) . 
\end{eqnarray}
Here, there is no central term classically contrast to the algebra for
the corresponding charges of (32). The center arise only as a quantum
mechanical 
effect of normal ordering [9, 11]. 

Now, the Dirac bracket algebra of $Q (f)$'s
becomes
\begin{eqnarray}
\{ Q (f), Q (g) \}^*\cong \{ Q_S (f), Q_{S} (g) \}^*
&\cong& Q_S ([f,g]_{Lie})  
\end{eqnarray}
and this is the same algebra of the Poisson bracket which is inferred
from the fact of $\{
Q_{B} (f), Q (g) \} \cong 0$. The gauge transformation which is
generated by $Q (f)$
and the realization of holography principle is the same as in Section
III. {\bf B}. 

\begin{center}
{\bf Acknowledgments}
\end{center}

I appreciate Profs. Roman Jackiw and Pillial Oh for stimulating discussions at
the early stage of this work and Profs. Choonkyu Lee, Jae Hyung Lee
and Hee Sung Song for hospitality of providing a financial support and 
researching facilities in Korea. I also thank Profs. M\'aximo Ba{\~n}ados, 
Klaus Bering, Steve Carlip, Antonio Garc\`ia, Yong-Wan Kim, Ian Kogan,
Juoko Mickelsson, Miguel 
Ortiz, and Young-Jai Park for helpful correspondence. This work was
supported in part by the Korea Science and Engineering Foundation
(KOSEF), Project No. 97-07-02-02-01-3 and 
KOSEF through the Center for Theoretical Physics at Seoul National
University.   

\newpage
\begin{center}
{\large \bf References}
\end{center}
\begin{description}

\item{[1]} E. Witten, Commun. Math. Phys. {121} (1989) 351; G. Moore
  and N. Seiberg, Phys. Lett. B220 (1989) 422; S. Elitzur, G. Moore,
  A. Schwimmer and N. Seiberg, Nucl. Phys. B326 (1989) 108. 

\item{[2]} S. Carlip, Phys. Rev. {D51} (1995) 632; Phys. Rev. D55
  (1997) 878.
\item{[3]} A. P. Balachandran, A. Momen and L. Chandar,
  Nucl. Phys. B461 (1996) 581; A. P. Balachandran, L. Chandar and
  E. Ercolessi, Int. J. Mod. Phys. A10 (1995) 1969; D. Bak, S. P. Kim,
  S. K. Kim, K.-S. Soh, and J. H. Lee, Phys. Rev. B58 (1998) 024002; A. P. 
  Balachandran {\it et ~al.}, e-print hep-th/9804145.  

\item{[4]} J. D. Brown and M. Henneaux, Commun. Math. Phys. {104}
  (1986) 207.

\item{[5]} M. Ba\~nados, C. Teitelboim and J. Zanelli, Phys. Rev. Lett. {69} 
(1992) 1849; M. Ba\~nados, M. Henneaux, C. Teitelboim and J. Zanelli, Phys. 
Rev. {D48} (1993) 1506.

\item{[6]} A. Strominger, e-print hep-th/9712251; D. Birmingham,
  I. Sachs and S. Sen, e-print hep-th/9801019.

\item{[7]} S. Deser, R. Jackiw, and Templeton, Phys. Rev. Lett. {
    48}, 975 (1982); Ann. Phys. (N.Y.) {140} (1982) 372; {185}
  (1988) 406 (E).

\item{[8]} A. P. Balachandran, G. Bimontz, K. S. Gupta and A. Stern, 
Int. J. Mod. Phys.  A 7 (1992) 4655; 
G. Bimonte, K. S. Gupta, and A. Stern, 
Int. J. Mod. Phys.  A 8 (1993) 653.

\item{[9]} T. Regge and C. Teitelboim, Ann. Phys. (N.Y.) 88 (1974)
  286; R. Benguria, P. Cordero, and C. Teitelboim, Nucl. Phys. B122
  (1977) 61; J. D. Brown and M. Henneaux, J. Math. Phys. 27 (1986) 489.

\item{[10]} M. Ba\~nados, Phys. Rev. {D52} (1996) 5816.
 
\item{[11]} S. Deser and R. Jackiw, Phys. Lett B139 (1984) 371;
  E. Witten, Commun. Math. Phys. 92 (1984) 455; L. D. Faddeev and R. Jackiw, 
Phys. Rev. Lett.  60 (1988) 1692; L. D. Faddeev, preprint HU-TFT-92-5.

\item{[12]} P. Oh and M.-I. Park, e-print hep-th/9805178.

\item{[13]} R. Jackiw, {\it in}  {Current Algebra and Anomalies} edited 
by S. B. Treiman, R. Jackiw, B. Zumino, and E. Witten (World Scientific, 
Singapore, 1985); D. Bak, R. Jackiw and S.-Y. Pi, 
Phys. Rev.  D 49 (1994) 6778.

\item{[14]} A. Ach\'ucaro and P. K. Townsend, Phys. Lett. {B180} (1986) 89; 
E. Witten, Nucl. Phys. {B311} (1988) 46.

\item{[15]} S. Deser and R. Jackiw, Ann. Phys. {153} (1984) 405;
E. Witten, Commun. Math. Phys. {137} (1991) 29.

\item{[16]} M. Ba\~nados, T. Brotz, and M. Ortiz, 
e-print hep-th/9802076; See also S. Carlip, e-print hep-th/9806026 for a
critical review.

\item{[17]} M.-I. Park, Phys. Lett. B (in press), e-print
  hep-th/9806119. 

\item{[18]} J. Maldacena and A. Strominger, J. High Energy Phys. 02
  (1998) 014, e-print gr-qc/9801096; M. Ba\~nados, T. Broz and
  M. Ortiz, e-print hep-th/9807216.

\item{[19]} P. A. M. Dirac, Lectures on quantum mechanics (Yeshiva
  University Press, New York 1964).

\item{[20]} M. Ba\~nados and A. Gomberoff, Phys. Rev. D55 (1997) 6162.

\item{[21]} O. Coussart, M. Henneaux and P. Driel,
  Class. Quant. Grav. 12 (1995) 2961; T. Lee, e-print hep-th/9805182; 9806113.

\item{[22]} M.-I. Park and Y.-J. Park, Phys. Rev. D58 R101702 (1998)
  (in press), hep-th/9803208.

\item{[23]} G.'t Hooft, Dimensional Reduction in Quantum Gravity {\it in}
   {Salam Fest} (World Scientific Co. Singapore, 1993);
  L. Susskind, J. Math. Phys. 36 (1995) 6377.

\item{[24]} E. Witten, e-print hep-th/9802150; L. Susskind and
  E. Witten, e-print hep-th/9805114.

\item{[25]} V. O. Soloviev, Phys. Lett. B292 (1992) 30.

\item{[26]} R. Jackiw, Phys. Rev. Lett. {41} (1979) 1635;
Acta. Phys. Austr. Suppl. {XXII} (1980) 383, 
reprinted {\it in}  {Diverse Topics in Theoretical and Mathematical Physics}
(World Scientific, Singapore, 1995).

\item{[27]} P. Goddard and D. Olive, Int. J. Mod. Phys. A1 (1986) 303;
  V. G. Knizhnik and A. B. Zamolochikov, Nucl. Phys. B247 (1984) 83;
  Ya I. Kogan and V. V. Fock, JETP Lett. 51 (1990) 210;
  Mod. Phys. Lett. A5 (1990) 1365; M. Bos and V. P. Nair,
  Int. J. Mod. Phys. A5 (1990) 959. 

\item{[28]} A. Chodos and C. B. Thorn, Nucl. Phys. B72 (1984) 509;
  E. D'Hoker and R. Jackiw, Phys. Rev. D26 (1982) 3517.

\item{[29]} J. Maldacena, hep-th/9711200.

\item{[30]} M. Henneaux and C. Teitelboim, {Quantization of Gauge
    Systems} (Princeton Univ. Press, Princeton, New Jersey, 1992).

\item{[31]} R. Sugano and T. Kimura, Phys. Rev. D41, 1247 (1990).

\item{[32]} J. A. Garc\`ia and J. M. Pons, Int.J. Mod. Phys A12 (1997) 451; 
{\it ibid.} 12 (1997) 451.

\item{[33]} J. Mickelsson, Lett. Math. Phys. {7} (1983) 45.

\item{[34]} G. V. Dunne and C. A. Trugenberger, Ann. Phys. (N.Y.) 204
  (1990) 281; Phys. Lett. B248 (1990) 311.


\item{[35]} M. Ba\~nados, K. Boutier, O. Cousaert, M. Henneaux, and M. Ortiz, 
e-print hep-th/9805165.

\item{[36]} J. Mickelsson, Commun. Math. Phys. {97} (1985) 361;
M. Ba\~nados, L. J. Garat and M. Henneaux, Nucl. Phys.
{B476} (1996) 611.

\item{[37]} S. Hyun, e-print hep-th/9704005.

\end{description}
\end{document}